%% file: main.tex
\RequirePackage{fix-cm}
\documentclass[runningheads]{llncs}

\newcommand{\rpm}{\sbox0{$1$}\sbox2{$\scriptstyle\pm$}
  \raise\dimexpr(\ht0-\ht2)/2\relax\box2 }

\usepackage{microtype}
\usepackage{epsfig} 
\usepackage{mathptmx} 
\usepackage{times} 
\usepackage{amsmath} 
\usepackage{amssymb}  
\usepackage{graphicx}
\usepackage{multirow}
\title{\LARGE \bf
Analysis of Big Data Technology for Health Care Services
}

\author{Dinesh Samuel Sathia Raj\inst{1}    \and
        Vijayakumar V\inst{1}                    \and
        Bharat Rawal\inst{2}                  \and
        Longzhi Yang\inst{3}             
}
\institute{ Vellore Institute of Technology, Inida
           \and
              Penn State University, USA
            \and
            Northumbria University, UK
}
\date{}
\begin{document}

\maketitle
\thispagestyle{empty}
\pagestyle{empty}

\begin{abstract}
Deep learning and other big data technologies have over time become very powerful and accurate. There are algorithms and models developed that have near human accuracy in their task. In health care, the amount of data available is massive and hence, these technologies have a great scope in health care. This paper reviews a few interesting contributions to the field specifically to medical imaging, genomics and patient health records.
\keywords{Medical Imaging \and Machine Learning \and Deep Learning \and Convolutional Neural Networks \and Recurrent Neural Networks \and LSTM \and Transfer Learning.}
\end{abstract}

\section{Introduction}
Big data technologies have found their way into almost every major domain and the health care domain is no exception. 
The pervasiveness of smart-wearables has lead to an explosive growth in health data and support. Products that harness the power of data science for medicine already exist. Apple's latest smart watch can detect heart attacks. Most WearOS smart watches can accurately detect the type of physical activity being carried out and hence recommend various activites. Omada Health, a company founded in 2011, has come up with a data driven system to help users reduce the risk of preventable diseases. MedAware is another startup that aims to eliminate prescription errors. Using data, their system learns over time and informs doctors about changes from the normal prescription for a particular case. 
\\This paper aims to discuss a few innovative advancements in important fields of health care. Image data in health care is found abundantly and many scans are images. Analysis of these images still mostly require a trained eye and computers are still not as accurate as trained professionals. Genomics or the study of genes is a field that deals with a lot of data. Each gene sequence has multiple permutations and finding patterns and connections among these genes has always been a challenge. Finally, electronic health records provide a massive source of temporal data about a patient and analysis of this data by a computer also proves to be very useful. Hence, this paper focuses on these three major domains of health care. 
\\The rest of the paper is structured as follows. Section II gives a brief overview of the various technologies being used in the medical domain. Section III discusses the various methods used in the analysis of image data. Section IV deals with deep learning in genomics. Section V deals with electronic health records(EHRs) and patient support techniques. Section VI concludes the paper.
\section{Technologies}
This section explains some of the most common data science techniques and ideas used in many of the papers surveyed in this work.
\subsection{Convolutional Neural Networks}
Convolutional Neural Networks(CNNs) are deep neural networks that are particularly efficient at analyzing images. These networks share weights within themselves and can hence detect shapes and features no matter where they appear. This also allows the network to learn much fewer parameters or weights compared to traditional neural networks.
In CNNs, features are learnt by convolving filters of various sizes with the input image. The convolved image is then passed on to deeper layers where the process is repeated. Deeper layers in a CNN tend to learn complex features while shallow layers may act as edge and corner detectors. 
\begin{figure}[ht]
    \centering
    \includegraphics[scale=0.3]{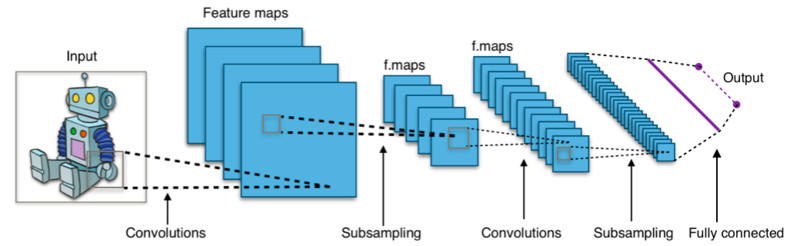}
    \caption{Typical CNN architecture with convolutions, pooling(subsampling) and fully connected layers.}
    \label{fig:cnn}
\end{figure}
Pooling layers, where neighbouring pixels are pooled using functions like $mean$ or $max$ further reduce the number of parameters to be learnt and also promote translational invariance.

\subsection{Sequence Models}
\begin{figure}[ht!]
    \centering
    \includegraphics[scale=0.75]{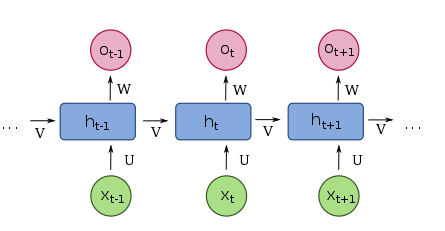}
    \caption{Pictorial representation of a simple RNN }
    \label{fig:rnn}
\end{figure}

With time dependent data like electrocardiograms (ECG) and electroencephalograms (EEG) making up a large part of medical data, recurrent neural networks play an important role in diagnosis. These networks have a temporal component and are hence very effective in analysing sequence data like speech recognition and translation. 
As seen in figure \ref{fig:rnn}, an RNN uses data learnt from the past inputs to predict outputs for the current timestep. 
One of the first effective RNN was the long short term memory(LSTM) network proposed by Hochreiter et al.\cite{lstm} in 1997. This network solved the issue of the network "forgetting" inputs received much before the current timestep. in 2014, Kyunghyun Cho et al.\cite{gru} proposed gated recurrent units(GRUs). Networks with GRUs are easier to work with and can be trained faster than LSTMs. 

\subsection{Support Vector Machines}
A support vector machine (SVM) is a supervised learning algorithm, proposed by Cortes et al.\cite{svm} in 1995, mainly used for two class classification. An SVM builds a hyper plane with the maximum margin to separate data. Assume data points from the first class are labelled 1 and those from the second class are labelled -1. Now, hyper-planes for these two classes can be defined by $W^Tx-b = 1$ for the class with label 1 and $W^Tx-b = -1$ for the class with label -1. The distance between these two classes can be found using the formula $\frac{2}{||w||}$. Hence, the aim of the classifier is now to minimize $||w||$ so that the distance between the hyper-planes is maximised. The hinge loss function \eqref{equation:one} is used to carry out this maximisation.
\begin{equation}
    \sum{max(0, 1 - y_i(W^Tx_i + b))}
    \label{equation:one}
\end{equation}
Adding regularisation and fine tuning the system, researchers came up with formula \eqref{equation:two}. The $\lambda$ term determines the trade-off between margin size and prediction accuracy and tuning this constant can reduce overfitting. 
\begin{equation}
    \frac{1}{n}[\sum^n_{i=1}{max(0, 1 - y_i(W^Tx_i + b))] + \lambda||W^2||}
    \label{equation:two}
\end{equation}
\subsection{Auto Encoders and Stacked Auto Encoders}
An auto encoder is a neural network that consists of three layers. The input layer, one hidden layer and an output layer. It is a method of unsupervised learning where the network learns a reduced representation of the input. This network works very well for dimensional reduction and is powerful in finding important patterns. While training, the task of the auto encoder is to find the best set of weights with which the input can be copied to the output. The size of the middle layer dictates how much dimensionality reduction is carried out. If the number of nodes in the input and hidden layers is the same, the weights learnt will essentially be linear since the output layer can now perfectly reconstruct the input image. The trade off between amount of reduction and quality of reconstruction is the major design goal of an auto encoder. In medical data, while dimensionality reduction is important, auto encoders are used to learn latent, important features. The auto encoders designed are forced to learn only the most important features of an image due to different constraints imposed on the network.
Stacked auto encoders are many auto encoders stacked one on top of the others. They follow a greedy approach while learning, as described by Benjio et al.\cite{sae}. The shallow layers of these networks learn more linear features while deeper layers are able to learn more complex features.

\section{Medical Imaging}
Images make up a large portion of medical data. From X-Rays to MRI scans, these images contain information that can be analysed through classification, detection and segmentation. Data science has provided the tools and ideas to automate these tasks and this section sheds light on few of the methods currently being used. 
\\One of the first uses of convolutional neural networks (CNNs) in the medical field was described way back in 1995 by Shih-Chung B. Lo et al.\cite{Lung}. They used CNNs to detect lung nodules. It took their model fifteen seconds to detect the presence of a nodule. An eternity in today's age. With advancements in both computational capacity and neural network architectures, the power and efficiency of image analysis will increase greatly. 
\\Medical data is not as freely available or as abundant as natural images. With vanilla CNNs requiring a lot of data to be efficient, researchers are forced to come up with new and improved architectures of CNNs. One of these improved architectures is the U-Net proposed by Ronneberger et al.\cite{unet}. The network is used for image segmentation and works better than traditional sliding window methods. The network has a down-scaling section as well as an up-scaling section. It essentially uses the down scaled features learned to construct the foreground. The network also has "skip connections" which mean the network is always aware about the whole image and data is not lost. Images can be segmented in a single forward pass through the network. U-Net won the ISIB EM segmentation challenge where they had to segment neural structures from electron microscopy slices in 2015. The U-Net architecture laid the path to other, more powerful networks with similar architectures like the V-Net proposed by Milletari et al.\cite{vnet} for example which worked even on 3D data.
\\Esteva A et al.\cite{skinCNN} describe using pre-trained weights to build more powerful system to detect skin diseases. Using transfer learning, they were able to build a system that can classify 757 diseases with an accuracy of 72.1$\pm$0.9\%. They trained the GoogLeNet architecture\cite{GoogLeNet}, pre-trained on 1.28 million images from 1000 classes, on a labelled dataset of 129,450 clinical images data and modified the output layer to predict one of the 757 diseases. Their method provided almost human like prediction for skin diseases. Yun Liu et al.\cite{transferCancer} also use transfer learning to detect Cancer metastases or spread of cancerous cells in breast cancer. They used Google's Inception v3 architecture and not only reported an accuracy of 92.4\% but also found two samples in the data set that had been labeled incorrectly.
\\While pre-trained weights seem to work well, recent research shows that it might not be as effective as described. Kaiming He et al.\cite{transferBad1} show that while using pre-trained weights might help speed up convergence, the accuracy of the network is no worse than training from scratch on even small datasets with around 10,000 images.They also show that using these pre-trained weights does not prevent over fitting except in a few cases. Medical images and day to day images like the ones in the ImageNet or COCO dataset are very distinguishable. Hence, using weights trained on these datasets might not be applicable to medical imaging. Maithra Raghu et al.\cite{transferBad2} in their paper show that using pre trained weights gives only minimal accuracy gain but increases the speed of convergence significantly. 
\\Kawahara et al.\cite{MultiResSkin} combined traditional CNNs but across multiple resolutions of the input image to classify skin lesions into 10 classes. Images of different resolutions are passed through similar neural networks and the final layer uses the weights of all the different CNNs to make a prediction. Their multi-tract CNN is able to capture interactions over multiple resolutions and hence report a higher accuracy than existing methods. Gao et al.\cite{CatMul} proposed a convolutional recursive neural network (CRNN) to carry out nuclear cataract grading. CRNNs learn representations of the different grades and then a support vector regressor uses these representations to predict the grade.
\\Kong et al.\cite{lstmCnn} proposed an interesting combination of CNNs and LSTMs to recognize end systole and end diastole frames in a cardiac sequence. The aim was to predict the frame in which the end systole and end diastole occur and they reported an average difference of 0.4 frames. The CNN encoded each frame while the LSTM brought in the temporal aspect. The method essentially used the coded CNN output as the input to the LSTM at every time step. This combination allowed the authors to overcome many of the issues that past models had. They also modelled a novel loss function to take into account the temporal aspect of the data.
\\Stacked auto encoders have been used by Benou et al.\cite{saeDenoise} to denoise Dynamic contrast-enhanced MRIs. Janowczyk et al.\cite{saeStain} proposed Stain Normalization using Sparse Auto Encoders(StaNoSA), a colour standardization method for digital histopathology. By using stacked  auto encoders, the authors were able to overcome the challenges posed by previous methods. These challenges included having a training set with similar tissue type representations. The auto encoder can identify the sub-type of the tissue in an unsupervised method. 
\\Convolutional Neural Networks are very widely used for almost all image related analysis. With powerful hardware and well documented software frameworks like tensorflow and caffe, there is a great scope for almost everyone to develop better models. With convolutions being the major operation when it comes to images, researchers are able to use different approaches to make those convolutions more meaningful and effective. Most models implement different loss functions or architectures with some combining different types of networks to include more data.
\section{Genetics and Genomics}
The availability of large datasets has led to the discovery of connections in genomics and inferences previously believed to be completely random. Contact prediction in proteins for example was thought to be rather random and independently distributed. Swark et al.\cite{prot} built a method to predict these contacts by using deep learning and also imposing specific constraints to improve accuracy. The authors do not use a traditional neural network but instead stack multiple random forest learners. By doing this, they were able to reduce the complexity, computational cost and training time of the model while preserving the ability to learn high level features.  
\\Precisely predicting transcriptional enhancers has been a major challenge due to the fact that these enhancers are highly cell type or tissue specific. As a result, building a single system to predict enhancers across multiple human cell types or tissues has been a major challenge. By predicting enhancers across multiple cell types, professionals can better understand the regulatory mechanism of these elements. Kleftogiannis et al.\cite{deep} proposed a model called DEEP to predict enhancers across multiple cell types be using an ensemble method and using data from multiple datasets, each dealing with a different representation of the enhancers. The authors combined three such models, DEEP-ENCODE, which specialises in predicting enhancers from the ENCODE repository; DEEP-FANTOM5, which specialises in predicting enhancers specifically expressed in particular organs and tissues; DEEP-VISTA model, which was trained on human in vivo-derived developmental enhancers. Multiple SVMs running the Gaussian kernel were then used to provide confidence scores to the input data. The DEEP-ENCODE component had 4000 classifiers, the DEEP-FANTOM5 had 50 classifiers and the DEEP-VISITA model had 10 classifiers. Finally, a neural network finally predicts if the sample is a candidate enhancer or not.  
\\The PEDLA framework proposed by Liu et al.\cite{pedla} further generalises the DEEP model and provides predictions for even more cell types. The authors claim PEDLA is able to learn the predictor from highly heterogeneous data and can also handle highly class imbalanced data in an unbiased way. PEDLA used the idea of using unsupervised learning and then fine tuning the network with supervised learning. Using iterative training, PEDLA learned features for the enhancers of one cell type and then used these weights as the initialisation for learning the features of the next class. If say there are 22 cell types, the model is trained 22 times, each subsequent model using the weights previously learned. This gave the model the flexibility of adding more cell types and increased the number of cell types where it could predict if the sample was an enhancer or not. 
\\Detection of cancer from gene expression poses a challenge due to the fact that the data used to make these predictions has enormous dimensions and is extremely complex. Deep learning methods can be used to tackle this problem and Danaee et al.\cite{cancer} proposed a model based on stacked denoising auto encoders along with SVMs and a shallow neural network to try and predict genes responsible for breast cancer. Due to the limited availability of training data for this task, the authors use synthetic minority over-sampling technique (SMOTE)\cite{smote} to create synthetic data. The model consists of an SAE used for dimensionality reduction followed by an SVM classifier and a shallow neural network. They also used principal component analysis for dimensionality reduction as a comparative study. The authors believe increasing the size of the dataset will lead to better training and hence a higher accuracy.
\\In genomics, data science is able to find patterns and connections that have been overlooked. Making sense of these connections still requires a specialist in genomics. The work carried out does seem promising and with more and more data being made available, the accuracy and impact big data technologies will have on genomics will only rise. 

\section{Patient Support}
Electric health records (EHRs) were initially used for archiving patient records, billing and other administrative tasks. With the rise of deep learning, researchers have found better use of this data. Deep patient, proposed by Miotto et al.\cite{deepPatient} uses EHRs to predict future disease and re-admittance of the patient. The authors use deep denoising auto encoders to convert the vast and varied data present in EHRs to a high order representation for each patient. This representation helps in predicting if a patient is likely to get another disease in the future. Even doctors notes from the EHRs have been used to ensure the maximum utilisation of the available data. Choi et al.\cite{ehrHF} use EHRs to predict heart failure in patients. Using natural language processing techniques like skip-gram\cite{skipgram} and GloVE\cite{glove}, both of which convert sentences to word vectors taking into consideration relationships between words. While training, data from 18 months prior to the detection of the heart failure is used for each patient. The representations developed from the EHRs are passed to a logistic regression classifier, an SVM, a neural network, and a K-nearest neighbours model for a comparative study. The neural network has the highest accuracy. With EHRs having historical data about patients, Choi et al.\cite{docAI} propose DoctorAI, a system that can predict diagnosis, medication and visit time using RNNs, specifically GRUs. Skip-gram is used here too to understand the notes and diagnosis provided by doctors. Deepr, proposed by Nguyen et al.\cite{deepr} predicts re-admittance within 3 to 6 months. The model converts text from the EHR into embeddings that can be placed in the euclidean space. Next, a convolutional layer receives input from a sliding window which passes chunks of words to it. Using convolutions here helps in finding "clinical motifs" or co-occurrences and patterns that may lead to re-admittance. 
\\As more and more hospitals adopt EHRs, more data is available to make algorithms more accurate. This explosion of data and technology could soon lead to mobile apps which can accurately diagnose and provide the initial treatment for many patients who may not be able to visit a real doctor.
\section{Discussion and Conclusion}
\input{ta.tex}
This paper shows the pervasiveness of big data technologies in the various important fields of health care. Table \ref{tab:table1} has a summary of all the methods described in this paper. We can see that neural networks are the most widely used method. Convolutional networks perform well on image data, LSTMs and GRUs perform well on data that has a temporal component and auto encoders create representations of complex data which can then be used by neural networks or SVMs to make predictions.
\\Looking at all the algorithms used, future research will most probably focus on building models that can learn predictions even on small datasets. Currently, CNNs and other neural network architectures require massive amounts of training data for accurate predictions. Some data in the health care domain is sensitive and hence not freely or widely available. Hence, algorithms which can perform well even on a small training set are important. Apart from this, we hope to see existing algorithms tweaked and tuned perfectly for specific application. Choosing the correct architecture, number of hidden layers and units, activation function etc all remain a challenge based on the application and this provides a large scope for new models and systems. Finally, solutions that combine multiple models or try to use cross domain knowledge will also be an area of focus for future research.
\\
\\
\\
\\
\\
\\
\\
\\
\\
\bibliographystyle{r.bst}

\bibliography{references}

\end{document}

%% file: ta.tex
\begin{table*}[t]
\centering
\caption{Summary of methods}
\label{tab:table1}
{\renewcommand{\arraystretch}{1.5}
\begin{tabular}{|l|p{5cm}|p{4cm}|p{3cm}|}
\hline
\multicolumn{1}{|c|}{\textbf{Domain}} & \multicolumn{1}{c|}{\textbf{Task}}                & \multicolumn{1}{c|}{\textbf{Method}}         & \multicolumn{1}{c|}{\textbf{Reference}}                  \\ \hline
                                      & Detection of Lung Nodules                         & CNN                                          & Shih-Chung et al. \cite{Lung}                            \\ \cline{2-4} 
                                      & Neural Structure Segmentation                     & U-Net                                        & Ronneberger et al. \cite{unet}                           \\ \cline{2-4} 
                                      & Segmentation of Prostate MRI Volumes              & V-Net                                        & Milletari et al. \cite{vnet}                             \\ \cline{2-4} 
                                      & Skin Disease Prediction                           & \multirow{2}{*}{Transfer Learning, CNN}      & Esteva A et. al. \cite{skinCNN}                          \\ \cline{2-2} \cline{4-4} 
Imaging                               & Detection of Cancer Metastases                    &                                              & Yun Liu et al. \cite{transferCancer}                     \\ \cline{2-4} 
                                      & Skin Lesion Classification                        & Multi Resolution CNN                         & Kawahara et al.\cite{MultiResSkin}                       \\ \cline{2-4} 
                                      & Nuclear Cataract Grading                          & Convolutional RNN                            & Gao et al. \cite{CatMul}                                 \\ \cline{2-4} 
                                      & End Systole and End Diastole prediction           & CNN+LSTM                                     & Kong et al. \cite{lstmCnn}                               \\ \cline{2-4} 
                                      & Colour Standardisation for digital histopathology & Stacked Auto Encoders                        & Janowczyk et al.\cite{saeStain}                          \\ \hline
                                      & Contact Prediction in Proteins                    & Stacked Random Forest Classifiers            & Swark et al. \cite{prot}                                 \\ \cline{2-4} 
Genomics                              & Transcriptional Enhancer Prediction               & DEEP, PEDLA                                  & Kleftogiannis et al. \cite{deep}; Liu et al.\cite{pedla} \\ \cline{2-4} 
                                      & Cancer Detection from Gene Expression             & Stacked Denoising Auto Encoders              & Danaee et al. \cite{cancer}                              \\ \hline
                                      & Prediction of Future Disease and Readmittance     & Stacked Denoising Auto Encoders              & Miotto et al. \cite{deepPatient}                         \\ \cline{2-4} 
EHR                                   & Heart Failure Prediction                          & Natural Language Processing + Neural Network & Choi et al. \cite{ehrHF}                                 \\ \cline{2-4} 
                                      & Predict Diagnosis, Medication and Visit Time      & RNN (GRU)                                    & Choi et al. \cite{docAI}                                 \\ \cline{2-4} 
                                      & Readmittance Prediction                           & CNN                                          & Nguyen et al. \cite{deepr}                               \\ \hline
\end{tabular}
}
\end{table*}